\documentclass[10pt]{article}
\usepackage{amsfonts,amsmath,amssymb,latexsym}
\usepackage{multirow}
\usepackage[longnamesfirst,round]{natbib}
\usepackage{hyperref,graphicx}
\usepackage{setspace}
\usepackage{enumitem}
\usepackage{booktabs}
\usepackage{color}
\newcommand{\e}{\mathbb{E}}
\newcommand{\var}{\mathbb{V}}
\newcommand{\p}{\mathbb{P}}
\newcommand{\bs}{\boldsymbol}
\newcommand{\mbf}{\mathbf}

\newcommand{\diag}{\mathop{\mathrm{diag}}}

\newcommand{\nrm}{\mathcal{N}}
\newcommand{\unif}{\mathcal{U}}

\newcommand{\bx}{\bs{x}}

\newcommand{\by}{\bs{y}}
\newcommand{\bY}{\bs{Y}}
\newcommand{\bz}{\bs{z}}
\newcommand{\bZ}{\bs{Z}}
\newcommand{\bU}{\bs{U}}

\newcommand{\bp}{\bs{p}}

\newcommand{\bzero}{\bs{0}}

\newcommand{\bbeta}{\bs{\beta}}

\newcommand{\bomega}{\bs{\omega}}

\newcommand{\bpsi}{\bs{\psi}}

\newcommand{\btheta}{\bs{\theta}}

\newcommand{\momega}{\mbf{\Omega}}

\newcommand{\id}{\mbf{I}}

\newcommand{\meat}{\bs{\mathcal{J}}}
\newcommand{\info}{\bs{\mathcal{I}}}

\newcommand{\cml}{\text{\textsc{cml}}}
\newcommand{\dt}{\text{\textsc{dt}}}

\newcommand{\bull}{\text{\raisebox{1pt}{\scalebox{.6}{$\bullet$}}}}

\title{Sklar's Omega: A Gaussian Copula-Based Framework for Assessing Agreement}
\author{John Hughes\,\thanks{jphughesjr@gmail.com}}

\date{}

\begin{document}

\maketitle

\begin{abstract}
The statistical measurement of agreement is important in a number of fields, e.g., content analysis, education, computational linguistics, biomedical imaging. We propose Sklar's Omega, a Gaussian copula-based framework for measuring intra-coder, inter-coder, and inter-method agreement as well as agreement relative to a gold standard. We demonstrate the efficacy and advantages of our approach by applying it to both simulated and experimentally observed datasets, including data from two medical imaging studies. Application of our proposed methodology is supported by our open-source R package, \texttt{sklarsomega}, which is available for download from the Comprehensive R Archive Network.\medskip

\end{abstract}

\section{Introduction} 
\label{intro}

We develop a model-based alternative to Krippendorff's $\alpha$ \citep{hayes2007answering}, a well-known nonparametric measure of agreement. In keeping with the naming convention that is evident in the literature on agreement (e.g., Spearman's $\rho$, Cohen's $\kappa$, Scott's $\pi$), we call our approach Sklar's $\omega$. Although Krippendorff's $\alpha$ is intuitive, flexible, and subsumes a number of other coefficients of agreement, we will argue that Sklar's $\omega$ improves upon $\alpha$ in (at least) the following ways. Sklar's $\omega$
\begin{itemize}
\item permits practitioners to simultaneously assess intra-coder agreement, inter-coder agreement, agreement with a gold standard, and, in the context of multiple scoring methods, inter-method agreement;
\item identifies the above mentioned types of agreement with intuitive, well-defined population parameters;
\item can accommodate any number of coders, any number of methods, any number of replications (per coder and/or per method), and missing values;
\item allows practitioners to use regression analysis to reveal important predictors of agreement (e.g., coder experience level, or time effects such as learning and fatigue);
\item provides complete inference, i.e., point estimation, interval estimation, diagnostics, model selection; and
\item performs more robustly in the presence of unusual coders, units, or scores.
\end{itemize}

The rest of this article is organized as follows. In Section~\ref{problem} we present an overview of the agreement problem, and state our assumptions. In Section~\ref{examples} we present three example applications of both Sklar's $\omega$ and Krippendorff's $\alpha$. These case studies showcase various advantages of our methodology. In Section~\ref{method} we specify the flexible, fully parametric statistical model upon which Sklar's $\omega$ is based. In Section~\ref{inference} we describe four approaches to frequentist inference for $\omega$, namely, maximum likelihood, distributional transform approximation, and composite marginal likelihood. We also consider a two-stage semiparametric method that first estimates the marginal distribution nonparametrically and then estimates the copula parameter(s) by conditional maximum likelihood. In Section~\ref{simulation} we use an extensive simulation study to assess the performance of Sklar's $\omega$ relative to Krippendorff's $\alpha$. In Section~\ref{package} we briefly describe our open-source R \citep{Ihak:Gent:r::1996} package, \texttt{sklarsomega}, which is available for download from the Comprehensive R Archive Network \citep{CRAN}. Finally, in Section~\ref{conclusion} we point out potential limitations of our methodology, and posit directions for future research on the statistical measurement of agreement.

\section{Measuring agreement}
\label{problem}

We feel it necessary to define the problem we aim to solve, for the literature on agreement contains two broad classes of methods. Methods in the first class seek to measure agreement while also explaining disagreement---by, for example, assuming differences among coders (as in \citet{aravind2017statistical}, for example). Although our approach permits one to use regression to explain systematic variation away from a gold standard, we are not, in general, interested in explaining disagreement. Our methodology is for measuring agreement, and so we do not typically accommodate (i.e., model) disagreement. For example, we assume that coders are exchangeable (unless multiple scoring methods are being considered, in which case we assume coder exchangeability within each method). This modeling orientation allows disagreement to count fully against agreement, as desired.

Although our understanding of the agreement problem aligns with that of Krippendorff's $\alpha$ and other related measures, we adopt a subtler interpretation of the results. According to \citet{krippendorff2012content}, social scientists often feel justified in relying on data for which agreement is at or above 0.8, drawing tentative conclusions from data for which agreement is at or above 2/3 but less than 0.8, and discarding data for which agreement is less than 2/3. We use the following interpretations instead (Table~\ref{tab:interpret}), and suggest---as do Krippendorff and others \citep{artstein2008inter,landiskoch}---that an appropriate reliability threshold may be context dependent.

\begin{table}[h]
\centering
\begin{tabular}{cl}
Range of Agreement & Interpretation\\\hline
$\phantom{0.2<\;}\omega\leq 0.2$ & Slight Agreement\\
$0.2<\omega\leq 0.4$ & Fair Agreement\\
$0.4<\omega\leq 0.6$ & Moderate Agreement\\
$0.6<\omega\leq 0.8$ & Substantial Agreement\\
$\phantom{0.2<\;}\omega>0.8$ & Near-Perfect Agreement\\
\end{tabular}
\caption{Guidelines for interpreting values of an agreement coefficient.}
\label{tab:interpret}
\end{table}

\section{Case studies}
\label{examples}

In this section we present three case studies that highlight some of the various ways in which Sklar's $\omega$ can improve upon Krippendorff's $\alpha$. The first example involves nominal data, the second example interval data, and the third example ordinal data.

\subsection{Nominal data analyzed previously by Krippendorff}

Consider the following data, which appear in \citep{krippendorff2013}. These are nominal values (in $\{1,\dots,5\}$) for twelve units and four coders. The dots represent missing values.

\begin{figure}[h]
   \centering
   \begin{tabular}{ccccccccccccc}
   & $u_1$ &  $u_2$ & $u_3$ & $u_4$ & $u_5$ & $u_6$ & $u_7$ & $u_8$ & $u_9$ & $u_{10}$ & $u_{11}$ & $u_{12}$\vspace{2ex}\\
   $c_1$ & 1 & 2 & 3 & 3 & 2 & 1 & 4 & 1 & 2 & \bull & \bull & \bull\\
   $c_2$ & 1 & 2 & 3 & 3 & 2 & 2 & 4 & 1 & 2 & 5 & \bull & 3\\
   $c_3$ & \bull & 3 & 3 & 3 & 2 & 3 & 4 & 2 & 2 & 5 & 1 & \bull\\
   $c_4$ & 1 & 2 & 3 & 3 & 2 & 4 & 4 & 1 & 2 & 5 & 1 & \bull
   \end{tabular}
   \caption{Some example nominal outcomes for twelve units and four coders, with a bit of missingness.}
   \label{fig:nominal}
\end{figure}

Note that all columns save the sixth are constant or nearly so. This suggests near-perfect agreement, yet a Krippendorff's $\alpha$ analysis of these data leads to a weaker conclusion. Specifically, using the discrete metric $d(x,y)=1\{x\neq y\}$ yields $\hat{\alpha}=0.74$ and bootstrap 95\% confidence interval (0.39, 1.00). (We used a bootstrap sample size of $n_b=$ 1,000, which yielded Monte Carlo standard errors (MCSE) \citep{Flegal:2008p1285} smaller than 0.001.) This point estimate indicates merely substantial agreement, and the interval implies that these data are consistent with agreement ranging from moderate to nearly perfect.

Our method produces $\hat{\omega}=0.89$ and $\omega\in(0.70, 0.98)$ ($n_b=$ 1,000; MCSEs $<$ 0.004), which indicate near-perfect agreement and at least substantial agreement, respectively. And our approach, being model based, furnishes us with estimated probabilities for the marginal categorical distribution of the response:
\[
\hat{\bp}=(\hat{p}_1,\hat{p}_2,\hat{p}_3,\hat{p}_4,\hat{p}_5)'=(0.25, 0.24, 0.23, 0.19, 0.09)'.
\]
Because we estimated $\omega$ and $\bp$ simultaneously, our estimate of $\bp$ differs substantially from the empirical probabilities, which are 0.22, 0.32, 0.27, 0.12, and 0.07, respectively.

The marked difference in these results can be attributed largely to the codes for the sixth unit. The relevant influence statistics are
\[
\delta_{\alpha}(\bull,-6)=\frac{\vert\hat{\alpha}_{\bull,-6}-\hat{\alpha}\vert}{\hat{\alpha}}=0.15
\]
and
\[
\delta_{\omega}(\bull,-6)=\frac{\vert\hat{\omega}_{\bull,-6}-\hat{\omega}\vert}{\hat{\omega}}=0.09,
\]
where the notation ``$\bull,-6$" indicates that all rows are retained and column 6 is left out. And so we see that column 6 exerts 2/3 more influence on $\hat{\alpha}$ than it does on $\hat{\omega}$. Since $\hat{\alpha}_{\bull,-6}=0.85$, inclusion of column 6 draws us away from what seems to be the correct conclusion for these data.

\subsection{Interval data from an imaging study of hip cartilage}

\begin{figure}[h]
   \centering
   \includegraphics[scale=.225]{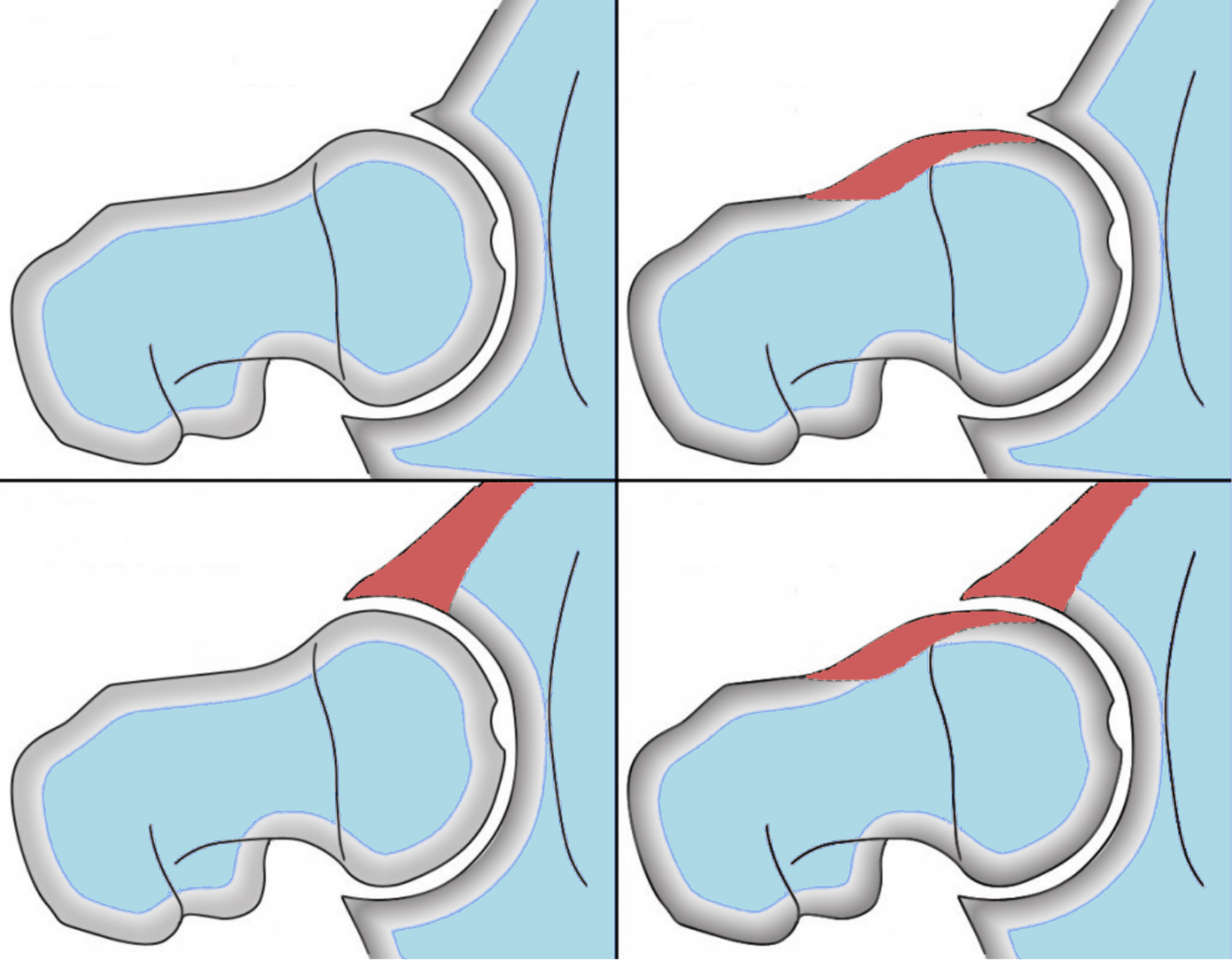}
   \caption{An illustration of femoroacetabular impingement (FAI). Top left: normal hip joint. Top right: cam type FAI. Bottom left: pincer type FAI. Bottom right: mixed type.}
   \label{fig:fai}
\end{figure}

\begin{figure}[h]
   \centering
   \includegraphics[scale=.45]{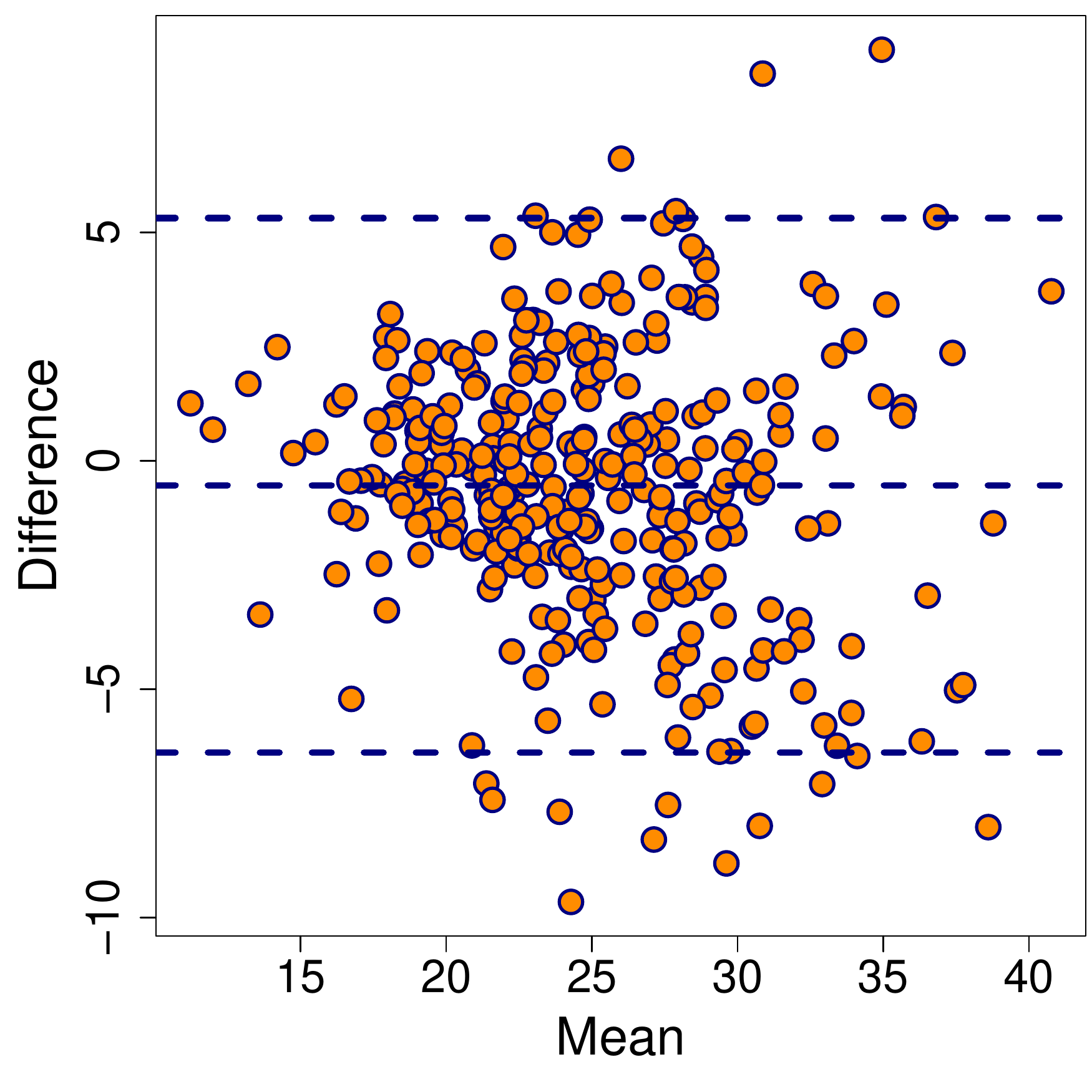}
   \caption{A Bland--Altman plot for the femoral cartilage data.}
   \label{fig:ba}
\end{figure}

The data for this example, some of which appear in Figure~\ref{fig:interval}, are 323 pairs of T2* relaxation times (a magnetic resonance quantity) for femoral cartilage \citep{nissi2015t2} in patients with femoroacetabular impingement (Figure~\ref{fig:fai}), a hip condition that can lead to osteoarthritis. One measurement was taken when a contrast agent was present in the tissue, and the other measurement was taken in the absence of the agent. The aim of the study was to determine whether raw and contrast-enhanced T2* measurements agree closely enough to be interchangeable for the purpose of quantitatively assessing cartilage health. The Bland--Altman plot \citep{altman1983measurement} in Figure~\ref{fig:ba} suggests good agreement: small bias, no trend, consistent variability.

\begin{figure}[h]
   \centering
   \begin{tabular}{cccccccccccc}
   & $u_1$ &  $u_2$ & $u_3$ & $u_4$ & $u_5$ & $\dots$ & $u_{319}$ & $u_{320}$ & $u_{321}$ & $u_{322}$ & $u_{323}$\vspace{2ex}\\
   $c_1$ & 27.3 & 28.5 & 29.1 & 31.2 & 33.0 & $\dots$ & 19.7 & 21.9 & 17.7 & 22.0 & 19.5\\
   $c_2$ & 27.8 & 25.9 & 19.5 & 27.8 & 26.6 & $\dots$ & 18.3 & 23.1 & 18.0 & 25.7 & 21.7
   \end{tabular}
   \caption{Raw and contrast-enhanced T2* values for femoral cartilage.}
   \label{fig:interval}
\end{figure}

We applied our procedure for each of three choices of parametric marginal distribution: Gaussian, Laplace, and Student's $t$ with noncentrality. The results are shown in Table~\ref{tab:femoral}, where the fourth and fifth columns give the estimated location and scale parameters for the three marginal distributions, the sixth column provides values of Akaike's information criterion (AIC) \citep{akaike1974new}, and the final column shows model probabilities \citep{burnham2011aic} relative to the $t$ model (since that model yielded the smallest value of AIC).

\begin{table}[h]
   \centering
   \resizebox{\textwidth}{!}{
   \begin{tabular}{rcccccc}
      \parbox[b]{1.5cm}{Marginal Model} & $\hat{\omega}$ & $\omega\in$ & Location & Scale & AIC & \parbox[b]{2cm}{Model\\Probability}\\\hline
      Gaussian & 0.837 & (0.803, 0.870) & 24.9 & \phantom{1}5.30 & 3,605 & 0.0002\\
      Laplace & 0.858 & (0.829, 0.886) & 24.0 & \phantom{1}4.34 & 3,643 & $\approx$ 0\phantom{0001111}\\
      $t$ & 0.862 & (0.833, 0.890) & 23.3 & 11.21 & 3,588 & 1\phantom{1111\,\,}\\
      Empirical & 0.846 & (0.808, 0.869) & $-$ & $-$ & $-$ & $-$\phantom{1111\,\,}
   \end{tabular}}
   \caption{Results from applying Sklar's $\omega$ to the femoral-cartilage data.}
   \label{tab:femoral}
\end{table}

\begin{figure}[h]
   \centering
   \includegraphics[scale=.45]{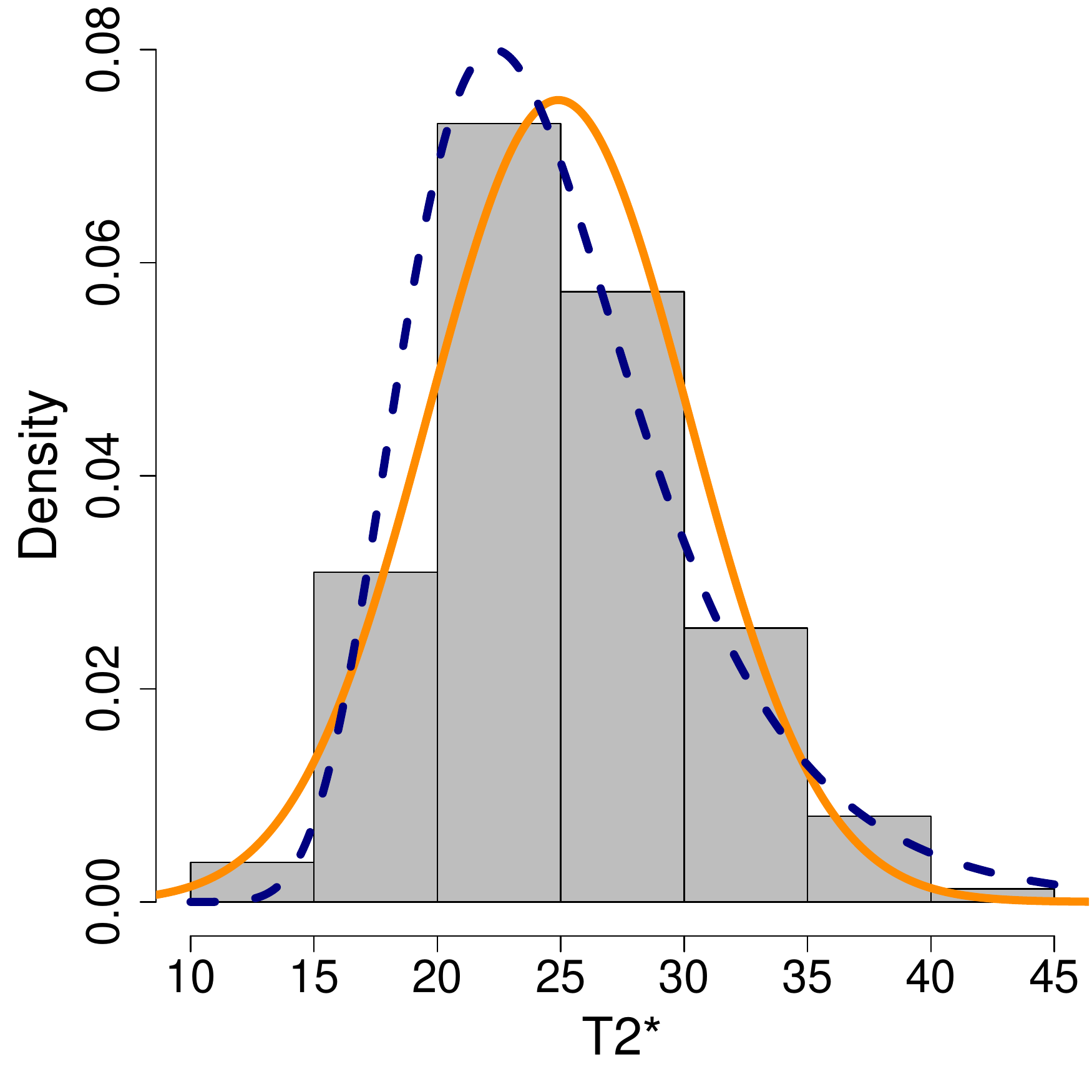}
   \caption{For the T2* data: histogram and fitted Gaussian and $t$ densities. The solid, orange curve is the fitted Gaussian density, and the dashed, blue curve is the fitted $t$ density.}
   \label{fig:femoral}
\end{figure}

We see that the estimates are comparable for the three choices of marginal distribution, yet the $t$ distribution is far superior to the others in terms of model probabilities. Figure~\ref{fig:femoral} provides visual corroboration: it is clear that the $t$ assumption proves more appropriate because it is able to capture the mild asymmetry of the marginal distribution. The $t$ assumption also yielded the largest estimate of $\omega$ and the narrowest confidence interval (arrived at using the method of maximum likelihood). In any case, we must conclude that there is near-perfect agreement between raw T2* and contrast-enhanced T2*.

Note that, for a sufficiently large sample, it may be advantageous to employ a nonparametric estimate of the marginal distribution (see Section~\ref{inference} for details). Results for this approach are shown in the final row of Table~\ref{tab:femoral}. We used a bootstrap sample size of 1,000 in computing the confidence interval. This yielded MCSEs smaller than 0.002.

A Krippendorff's $\alpha$ analysis gave $\hat{\alpha}=0.837$ and $\alpha\in(0.802, 0.864)$ ($n_b=$ 1,000; MSCEs $\approx$ 0.001). Since $\alpha$ implicitly assumes Gaussianity---$\alpha$ is the intraclass correlation for the ordinary one-way mixed-effects ANOVA model, and so $\hat{\alpha}$ is a ratio of sums of squares \citep{artstein2008inter}---it is not surprising that the $\alpha$ results are quite similar to the results obtained by Sklar's $\omega$ with Gaussian marginals.

\subsection{Ordinal data from an imaging study of liver herniation}

The data for this example, some of which are shown in Figure~\ref{fig:ordinal}, are liver-herniation scores (in $\{1,\dots,5\}$) assigned by two coders (radiologists) to magnetic resonance images (MRI) of the liver in a study pertaining to congenital diaphragmatic hernia (CDH) \citep{Danull}. The five grades are described in Table~\ref{tab:grades}.

\begin{figure}[h]
   \centering
   \begin{tabular}{cccccccccccc}
   & $u_1$ &  $u_2$ & $u_3$ & $u_4$ & $u_5$ & $\dots$ & $u_{43}$ & $u_{44}$ & $u_{45}$ & $u_{46}$ & $u_{47}$\vspace{2ex}\\
   $c_{11}$ & 2 & 4 & 4 & 4 & 4 & $\dots$ & 2 & 1 & 2 & 1 & 1\\
   $c_{12}$ & 2 & 4 & 5 & 4 & 4 & $\dots$ & 2 & 1 & 2 & 1 & 1\\
   $c_{21}$ & 3 & 5 & 5 & 5 & 4 & $\dots$ & 2 & 2 & 2 & 1 & 1\\
   $c_{22}$ & 3 & 5 & 5 & 4 & 4 & $\dots$ & 2 & 2 & 2 & 1 & 1
   \end{tabular}
   \caption{Ordinal scores for MR images of the liver. Each coder scored each unit twice.}
   \label{fig:ordinal}
\end{figure}

\begin{table}[h]
   \centering
   \begin{tabular}{cl}
   Grade & Description\\\hline
   1  & No herniation of liver into the fetal chest\\
2 & Less than half of the ipsilateral thorax is occupied by the fetal liver\\
3 & Greater than half of the thorax is occupied by the fetal liver\\
4 & The liver dome reaches the thoracic apex\\
5 & \parbox[t]{10cm}{The liver dome not only reaches the thoracic apex but also extends\\ across the thoracic midline}
   \end{tabular}
   \caption{Liver-herniation grades for the CDH study.}
   \label{tab:grades}
\end{table}

Each coder scored each of the 47 images twice, and so we are interested in assessing both intra-coder and inter-coder agreement. We can accomplish both goals with a single analysis by choosing an appropriate form for our copula correlation matrix $\momega$. Specifically, we let $\momega$ be block diagonal: $\momega=\diag(\momega_i)$, where the block for unit $i$ is given by
\[
\momega_i=
\bordermatrix{ & c_{11} & c_{12} & c_{21} & c_{22}\cr
c_{11} & 1 & \omega_1  & \omega_{12}  & \omega_{12}\cr
c_{12} & \omega_1 &  1 &  \omega_{12} & \omega_{12}\cr
c_{21} & \omega_{12} & \omega_{12}  & 1  & \omega_2\cr
c_{22} & \omega_{12} &  \omega_{12} & \omega_2  & 1
},
\]
$\omega_1$ being the intra-coder agreement for the first coder, $\omega_2$ being the intra-coder agreement for the second coder, and $\omega_{12}$ being the inter-coder agreement. See Section~\ref{method} for more information regarding useful correlation structures for agreement.

Our results (from a single analysis), and Krippendorff's $\alpha$ results (three separate analyses), are shown in Table~\ref{tab:liver}. We see that the point estimates for the two approaches are comparable (since the outcomes are approximately Gaussian). Our method produced considerably wider intervals, however. This is not surprising given that Sklar's $\omega$ estimates all three correlation parameters simultaneously and accounts for our uncertainty regarding the marginal probabilities. By contrast, Krippendorff's $\alpha$ can yield optimistic inference, as we show by simulation in Section~\ref{simulation}.

\begin{table}[h]
   \centering
   \resizebox{\textwidth}{!}{
   \begin{tabular}{rrc}
      Method & Agreement & Interval (MCSEs)\\\hline
      $\alpha$ (Intra-Agreement for Coder 1) & $\hat{\alpha}_1=0.987$ & (0.957, 1.000) ($\approx$ 0.002)\\
      $\alpha$ (Intra-Agreement for Coder 2) & $\hat{\alpha}_2=0.988$ & (0.958, 1.000) ($\approx$ 0.001)\\
      $\alpha$ (Inter-Agreement) & $\hat{\alpha}_{12}=0.956$ & (0.917, 0.980) ($\leq$ 0.002)\\\hline
      $\omega$ (Intra-Agreement for Coder 1) & $\hat{\omega}_1=0.987$ & (0.897, 0.995) ($\leq$ 0.002)\\
      $\omega$ (Intra-Agreement for Coder 2) & $\hat{\omega}_2=0.991$ & (0.835, 0.994) ($\leq$ 0.005)\\
      $\omega$ (Inter-Agreement) & $\hat{\omega}_{12}=0.965$ & (0.838, 0.996) ($\leq$ 0.006)
   \end{tabular}}
   \caption{Results from applying Krippendorff's $\alpha$ and Sklar's $\omega$ to the liver scores. The intervals are 95\% bootstrap intervals, where the bootstrap sample size was $n_b=$ 1,000.}
   \label{tab:liver}
\end{table}

\section{Our model}
\label{method}

The statistical model underpinning Sklar's $\omega$ is a Gaussian copula model \citep{xue2000multivariate}. We begin by specifying the model in full generality. Then we consider special cases of the model that speak to the tasks listed in Section~\ref{intro} and the assumptions and levels of measurement presented in Section~\ref{problem}.

The stochastic form of the Gaussian copula model is given by
\begin{align}
\label{gausscop}
\nonumber\bZ = (Z_1,\dots,Z_n)'  & \; \sim\;  \nrm(\bzero,\momega)\\
\nonumber U_i = \Phi(Z_i) & \;\sim\; \unif(0,1)\;\;\;\;\;\;\;(i=1,\dots,n)\\
Y_i = F_i^{-1}(U_i) & \;\sim\; F_i,
\end{align}
where $\momega$ is a correlation matrix, $\Phi$ is the standard Gaussian cdf, and $F_i$ is the cdf for the $i$th outcome $Y_i$. Note that $\bU=(U_1,\dots, U_n)'$ is a realization of the Gaussian copula, which is to say that the $U_i$ are marginally standard uniform and exhibit the Gaussian correlation structure defined by $\momega$. Since $U_i$ is standard uniform, applying the inverse probability integral transform to $U_i$ produces outcome $Y_i$ having the desired marginal distribution $F_i$.

In the form of Sklar's $\omega$ that most closely resembles Krippendorff's $\alpha$, we assume that all of the outcomes share the same marginal distribution $F$. The choice of $F$ is then determined by the level of measurement. While Krippendorff's $\alpha$ typically employs two different metrics for nominal and ordinal outcomes, we assume the categorical distribution
\begin{align}
\label{cat}
\nonumber p_k &= \p(Y=k)\;\;\;\;\;\;(k=1,\dots,K)\\
\sum_k p_k &= 1
\end{align}
for both levels of measurement, where $K$ is the number of categories. For $K=2$, (\ref{cat}) is of course the Bernoulli distribution.

Note that when the marginal distributions are discrete (in our case, categorical), the joint distribution corresponding to (\ref{gausscop}) is uniquely defined only on the support of the marginals, and the dependence between a pair of random variables depends on the marginal distributions as well as on the copula. \citet{genest2007primer} described the implications of this and warned that, for discrete data, ``modeling and interpreting dependence through copulas is subject to caution." But \citeauthor{genest2007primer} go on to say that copula parameters may still be interpreted as dependence parameters, and estimation of copula parameters is often possible using fully parametric methods. It is precisely such methods that we recommend in Section~\ref{inference}, and evaluate through simulation in Section~\ref{simulation}.

For interval outcomes $F$ can be practically any continuous distribution. Our R package supports the Gaussian, Laplace, Student's $t$, and gamma distributions. The Laplace and $t$ distributions are useful for accommodating heavier-than-Gaussian tails, and the $t$ and gamma distributions can accommodate asymmetry. Perhaps the reader can envision more ``exotic" possibilities such as mixture distributions (to handle multimodality or excess zeros, for example).

Another possibility for continuous outcomes is to first estimate $F$ nonparametrically, and then estimate the copula parameters in a second stage. In Section~\ref{inference} we will provide details regarding this approach.

Finally, the natural choice for ratio outcomes is the beta distribution, the two-parameter version of which is supported by package \texttt{sklarsomega}.

Now we turn to the copula correlation matrix $\momega$, the form of which is determined by the question(s) we seek to answer. If we wish to measure only inter-coder agreement, as is the case for Krippendorff's $\alpha$, our copula correlation matrix has a very simple structure: block diagonal, where the $i$th block corresponds to the $i$th unit $(i=1,\dots,n_u)$ and has a compound symmetry  structure. That is,
\[
\momega = \diag(\momega_i),
\]
where
\[
\momega_i = \bordermatrix{ & c_1 & c_2 & \dots & c_{n_c} \cr
c_1 & 1 & \omega  &\dots & \omega\cr
c_2 & \omega &  1 &  \dots & \omega\cr
\vdots & \vdots & \vdots  & \ddots  & \vdots\cr
c_{n_c} & \omega &  \omega & \dots  & 1
}
\]

On the scale of the outcomes, $\omega$'s interpretation depends on the marginal distribution. If the outcomes are Gaussian, $\omega$ is the Pearson correlation between $Y_{ij}$ and $Y_{ij'}$, and so the outcomes carry exactly the correlation structure codified in $\momega$. If the outcomes are non-Gaussian, the interpretation of $\omega$ (still on the scale of the outcomes) is more complicated. For example, if the outcomes are Bernoulli, $\omega$ is often called the tetrachoric correlation between those outcomes. Tetrachoric correlation is constrained by the marginal distributions. Specifically, the maximum correlation for two binary random variables is
\[
\min\left\{\sqrt{\frac{p_1(1-p_2)}{p_2(1-p_1)}},\sqrt{\frac{p_2(1-p_1)}{p_1(1-p_2)}}\right\},
\]
where $p_1$ and $p_2$ are the expectations \citep{prentice1988correlated}. More generally, the marginal distributions impose bounds, called the Fr\'{e}chet--Hoeffding bounds, on the achievable correlation \citep{Nelsen2006An-Introduction}. For most scenarios, the Fr\'{e}chet--Hoeffding bounds do not pose a problem for Sklar's $\omega$ because we typically assume that our outcomes are identically distributed, in which case the bounds are $-1$ and 1. (We do, however, impose our own lower bound of 0 on $\omega$ since we aim to measure agreement.)

In any case, $\omega$ has a uniform and intuitive interpretation for suitably transformed outcomes, irrespective of the marginal distribution. Specifically,
\[
\omega=\rho\left[\Phi^{-1}\{F(Y_{ij})\},\,\Phi^{-1}\{F(Y_{ij'})\}\right],
\]
where $\rho$ denotes Pearson's correlation and the second subscripts index the scores within the $i$th unit ($j,j'\in\{1,\dots,n_c\}:j\neq j'$). 

By changing the structure of the blocks $\momega_i$ we can use Sklar's $\omega$ to measure not only inter-coder agreement but also a number of other types of agreement. For example, should we wish to measure agreement with a gold standard, we might employ
\[
\momega_i = \bordermatrix{ & g & c_1 & c_2 & \dots & c_{n_c} \cr
g & 1 & \omega_g  & \omega_g & \dots & \omega_g\cr
c_1 & \omega_g &  1 &  \omega_c & \dots & \omega_c\cr
c_2 & \omega_g & \omega_c & 1 & \dots & \omega_c\cr
\vdots & \vdots & \vdots  & \vdots & \ddots  & \vdots\cr
c_{n_c} & \omega_g &  \omega_c & \omega_c & \dots  & 1
}.
\]
In this scheme $\omega_g$ captures agreement with the gold standard, and $\omega_c$ captures inter-coder agreement.

In a more elaborate form of this scenario, we could include a regression component in an attempt to identify important predictors of agreement with the gold standard. This could be accomplished by using a cdf to link coder-specific covariates with $\omega_g$. Then the blocks in $\momega$ might look like
\[
\momega_i = \bordermatrix{ & g & c_1 & c_2 & \dots & c_{n_c} \cr
g & 1 & \omega_{g1}  & \omega_{g2} & \dots & \omega_{gn_c}\cr
c_1 & \omega_{g1} &  1 &  \omega_c & \dots & \omega_c\cr
c_2 & \omega_{g2} & \omega_c & 1 & \dots & \omega_c\cr
\vdots & \vdots & \vdots  & \vdots & \ddots  & \vdots\cr
c_{n_c} & \omega_{gn_c} &  \omega_c & \omega_c & \dots  & 1
},
\]
where $\omega_{gj}=H(\bx_j'\bbeta)$, $H$ being a cdf, $\bx_j$ being a vector of covariates for coder $j$, and $\bbeta$ being regression coefficients.

For our final example we consider a complex study involving a gold standard, multiple scoring methods, multiple coders, and multiple scores per coder. In the interest of concision, suppose we have two methods, two coders per method, two scores per coder for each method, and gold standard measurements for the first method. Then $\momega_i$ is given by
\[
\momega_i = \bordermatrix{ & g_1 & c_{111} & c_{112} & c_{121} & c_{122} & c_{211} & c_{212} & c_{221} & c_{222}\cr
g_1 & 1 & \omega_{g1}  & \omega_{g1} & \omega_{g1} & \omega_{g1} & 0 & 0 & 0 & 0\cr
c_{111} & \omega_{g1} &  1 &  \omega_{11\bull} & \omega_{1\bull\bull} & \omega_{1\bull\bull} & \omega_{\bull\bull\bull} & \omega_{\bull\bull\bull} & \omega_{\bull\bull\bull} & \omega_{\bull\bull\bull}\cr
c_{112} & \omega_{g1} & \omega_{11\bull} & 1 & \omega_{1\bull\bull} & \omega_{1\bull\bull} & \omega_{\bull\bull\bull} & \omega_{\bull\bull\bull} & \omega_{\bull\bull\bull} & \omega_{\bull\bull\bull} &\cr
c_{121} & \omega_{g1} & \omega_{1\bull\bull} & \omega_{1\bull\bull} & 1 & \omega_{12\bull} & \omega_{\bull\bull\bull} & \omega_{\bull\bull\bull} & \omega_{\bull\bull\bull} & \omega_{\bull\bull\bull} &\cr
c_{122} & \omega_{g1} & \omega_{1\bull\bull} & \omega_{1\bull\bull} & \omega_{12\bull} & 1 & \omega_{\bull\bull\bull} & \omega_{\bull\bull\bull} & \omega_{\bull\bull\bull} & \omega_{\bull\bull\bull} &\cr
c_{211} & 0 & \omega_{\bull\bull\bull} & \omega_{\bull\bull\bull} & \omega_{\bull\bull\bull} & \omega_{\bull\bull\bull} &  1 & \omega_{21\bull} & \omega_{2\bull\bull} & \omega_{2\bull\bull}\cr
c_{212} & 0 & \omega_{\bull\bull\bull} & \omega_{\bull\bull\bull} & \omega_{\bull\bull\bull} & \omega_{\bull\bull\bull} & \omega_{21\bull} & 1 & \omega_{2\bull\bull} & \omega_{2\bull\bull}\cr
c_{221} & 0 & \omega_{\bull\bull\bull} & \omega_{\bull\bull\bull} & \omega_{\bull\bull\bull} & \omega_{\bull\bull\bull} & \omega_{2\bull\bull} & \omega_{2\bull\bull} & 1 & \omega_{22\bull}\cr
c_{222} & 0 & \omega_{\bull\bull\bull} & \omega_{\bull\bull\bull} & \omega_{\bull\bull\bull} & \omega_{\bull\bull\bull} & \omega_{2\bull\bull} & \omega_{2\bull\bull} & \omega_{22\bull} & 1\cr
},
\]
where the subscript $mcs$ denotes score $s$ for coder $c$ of method $m$. Thus $\omega_{g1}$ represents agreement with the gold standard for the first method, $\omega_{11\bull}$ represents intra-coder agreement for the first coder of the first method, $\omega_{12\bull}$ represents intra-coder agreement for the second coder of the first method, $\omega_{1\bull\bull}$ represents inter-coder agreement for the first method, and so on, with $\omega_{\bull\bull\bull}$ representing inter-method agreement.

Note that, for a study involving multiple methods, it may be reasonable to assume a different marginal distribution for each method. In this case, the Fr\'{e}chet--Hoeffding bounds may be relevant, and, if some marginal distributions are continuous and some are discrete, maximum likelihood inference is infeasible (see the next section for details).

\section{Approaches to inference for $\omega$}
\label{inference}

When the response is continuous, i.e., when the level of measurement is interval or ratio, we recommend maximum likelihood (ML) inference for Sklar's $\omega$. When the marginal distribution is a categorical distribution (for nominal or ordinal level of measurement), maximum likelihood inference is infeasible because the log-likelihood, having $\Theta(2^n)$ terms, is intractable for most datasets. In this case we recommend the distributional transform (DT) approximation or composite marginal likelihood (CML), depending on the number of categories and perhaps the sample size. If the response is binary, composite marginal likelihood is indicated even for large samples since the DT approach tends to perform poorly for binary data. If there are three or four categories, the DT approach may perform at least adequately for larger samples, but we still favor CML for such data. For five or more categories, the DT approach performs well and yields a more accurate estimator than does the CML approach. The DT approach is also more computationally efficient than the CML approach.

\subsection{The method of maximum likelihood for Sklar's $\omega$}

For correlation matrix $\momega(\bomega)$, marginal cdf $F(y\mid\bpsi)$, and marginal pdf $f(y\mid\bpsi)$, the log-likelihood of the parameters $\btheta=(\bomega',\bpsi')'$ given observations $\by$ is
\begin{align}
\label{loglik}
\ell_\textsc{ml}(\btheta\mid\by)=-\frac{1}{2}\log\vert\momega\vert-\frac{1}{2}\bz'(\momega^{-1}-\id)\bz+\sum_i\log f(y_i),
\end{align}
where $z_i=\Phi^{-1}\{F(y_i)\}$ and $\id$ denotes the $n\times n$ identity matrix. We obtain $\hat{\btheta}_\textsc{ml}$ by minimizing $-\ell_\textsc{ml}$. For all three approaches to inference---ML, DT, CML---we use the optimization algorithm proposed by \citet{byrd1995limited} so that $\bomega$, and perhaps some elements of $\bpsi$, can be appropriately constrained. To estimate an asymptotic confidence ellipsoid we of course use the observed Fisher information matrix, i.e., the estimate of the Hessian matrix at $\hat{\btheta}_\textsc{ml}$:
\[
\{\btheta:(\hat{\btheta}_\textsc{ml}-\btheta)'\,\hat{\info}_\textsc{ml}\,(\hat{\btheta}_\textsc{ml}-\btheta)\leq\chi^2_{1-\alpha,q}\},
\]
where $\hat{\info}_\textsc{ml}$ denotes the observed information and $q=\dim(\btheta)$.

Optimization of $\ell_\textsc{ml}$ is insensitive to the starting value for $\bomega$, but it can be important to choose an initial value $\bpsi_0$ for $\bpsi$ carefully. For example, if the assumed marginal family is $t$, we recommend $\bpsi_0=(\mu_0,\nu_0)'=(\text{med}_n,\text{mad}_n)'$ \citep{Serfling:2009p1313}, where $\mu$ is the noncentrality parameter, $\nu$ is the degrees of freedom, $\text{med}_n$ is the sample median, and $\text{mad}_n$ is the sample median absolute deviation from the median. For the Gaussian and Laplace distributions we use the sample mean and standard deviation. For the gamma distribution we recommend $\bpsi_0=(\alpha_0,\beta_0)'$, where
\begin{align*}
\alpha_0 &= \bar{Y}^2/S^2\\
\beta_0 &= \bar{Y}/S^2,
\end{align*}
for sample mean $\bar{Y}$ and sample variance $S^2$. Similarly, we provide initial values
\begin{align*}
\alpha_0 &= \bar{Y}\left\{\frac{\bar{Y}(1-\bar{Y})}{S^2}-1\right\}\\
\beta_0 &= (1-\bar{Y})\left\{\frac{\bar{Y}(1-\bar{Y})}{S^2}-1\right\}
\end{align*}
when the marginal distribution is beta. Finally, for a categorical distribution we use the empirical probabilities.

\subsection{The distributional transform method}

When the marginal distribution is discrete (in our case, categorical), the log-likelihood does not have the simple form given above because $z_i=\Phi^{-1}\{F(y_i)\}$ is not standard Gaussian (since $F(y_i)$ is not standard uniform if $F$ has jumps). In this case the true log-likelihood is intractable unless the sample is rather small. An appealing alternative to the true log-likelihood is an approximation based on the distributional transform.

It is well known that if $Y\sim F$ is continuous, $F(Y)$ has a standard uniform distribution. But if $Y$ is discrete, $F(Y)$ tends to be stochastically larger, and $F(Y^-)=\lim_{x\nearrow Y}F(x)$ tends to be stochastically smaller, than a standard uniform random variable. This can be remedied by stochastically ``smoothing" $F$'s discontinuities. This technique goes at least as far back as \citet{Ferguson:1969p1279}, who used it in connection with hypothesis tests. More recently, the distributional transform has been applied in a number of other settings---see, e.g., \citet{ruschendorf1981stochastically}, \citet{burgert2006optimal}, and \citet{Ruschendorf:2009p1281}.

Let $W\sim\mathcal{U}(0,1)$, and suppose that $Y\sim F$ and is independent of $W$. Then the distributional transform
\[
G(W,Y)=WF(Y^-)+(1-W)F(Y)
\]
follows a standard uniform distribution, and $F^{-1}\{G(W,Y)\}$ follows the same distribution as $Y$.

\citet{Kazianka:2010p941} suggested approximating $G(W,Y)$ by replacing it with its expectation with respect to $W$:
\begin{align*}
G(W,Y) &\approx \e_W G(W,Y)\\
&= \e_W\{WF(Y^-)+(1-W)F(Y)\}\\
&= \e_W WF(Y^-) + \e_W (1-W)F(Y)\\
&= F(Y^-)\e_W W + F(Y)\e_W(1-W)\\
&= \frac{F(Y^-) + F(Y)}{2}.
\end{align*}
To construct the approximate log-likelihood for Sklar's $\omega$, we replace $F(y_i)$ in (\ref{loglik}) with
\[
\frac{F(y_i^-) + F(y_i)}{2}.
\]
If the distribution has integer support, this becomes
\[
\frac{F(y_i-1) + F(y_i)}{2}.
\]

This approximation is crude, but it performs well as long as the discrete distribution in question has a sufficiently large variance \citep{Kazianka2013}. For Sklar's $\omega$, we recommend using the DT approach when the scores fall into five or more categories.

Since the DT-based objective function is misspecified, using $\hat{\info}_\textsc{dt}$ alone leads to optimistic inference unless the number of categories is large. This can be overcome by using a sandwich estimator \citep{godambe1960optimum} or by doing a bootstrap \citep{davison1997bootstrap}.

\subsection{Composite marginal likelihood}

For nominal or ordinal outcomes falling into a small number of categories, we recommend a composite marginal likelihood \citep{Lindsay:1988p1155,varin2008composite} approach to inference. Our objective function comprises pairwise likelihoods (which implies the assumption that any two pairs of outcomes are independent). Specifically, we work with log composite likelihood
\begin{align*}
\ell_\cml(\btheta\mid\by) &= \mathop{\mathop{\sum_{i\in\{1,\dots,n-1\}}}_{j\in\{i+1,\dots,n\}}}_{\momega_{ij}\neq 0}\log\left\{\sum_{j_1=0}^1\sum_{j_2=0}^1(-1)^k\Phi_{\momega^{ij}}(z_{ij_1},z_{jj_2})\right\},
\end{align*}
where $k=j_1+j_2$, $\Phi_{\momega^{ij}}$ denotes the cdf for the bivariate Gaussian distribution with mean zero and correlation matrix
\[
\momega^{ij}=\begin{pmatrix}1&\momega_{ij}\\\momega_{ij}&1\end{pmatrix},
\]
$z_{\bull 0}=\Phi^{-1}\{F(y_\bull)\}$, and $z_{\bull 1}=\Phi^{-1}\{F(y_\bull-1)\}$. Since this objective function, too, is misspecified, bootstrapping or sandwich estimation is necessary.

\subsection{Sandwich estimation for the DT and CML procedures}
\label{sandwich}

As we mentioned above, the DT and CML objective functions are misspecified, and so the asymptotic covariance matrices of $\hat{\btheta}_\dt$ and $\hat{\btheta}_\cml$ have sandwich forms \citep{godambe1960optimum,Geyer2005Le-Cam-Made-Sim}. Specifically, we have
\begin{align*}
\sqrt{n}(\hat{\btheta}_\cml-\btheta) &\;\;\Rightarrow_n\;\; \nrm\{\bzero,\;\info_\cml^{-1}(\btheta)\meat_\cml(\btheta)\info_\cml^{-1}(\btheta)\}\\
\sqrt{n}(\hat{\btheta}_\dt-\btheta) &\;\;\Rightarrow_n\;\; \nrm\{\bzero,\;\info_\dt^{-1}(\btheta)\meat_\dt(\btheta)\info_\dt^{-1}(\btheta)\},
\end{align*}
where $\info_\bull$ is the appropriate Fisher information matrix and $\meat_\bull$ is the variance of the score:
\[
\meat_\bull(\btheta)=\var\nabla\ell_\bull(\btheta\mid\bY).
\]
We recommend that $\meat_\bull$ be estimated using a parametric bootstrap, i.e, our estimator of $\meat_\bull$ is
\[
\hat{\meat}_\bull(\btheta)=\frac{1}{n_b}\sum_{j=1}^{n_b}\nabla\nabla^\prime\ell_\bull(\hat{\btheta}_\bull\mid\bY^{(j)}),
\]
where $n_b$ is the bootstrap sample size and the $\bY^{(j)}$ are datasets simulated from our model at $\btheta=\hat{\btheta}_\bull$. This approach performs well, as our simulation results show, and is considerably more efficient computationally than a ``full" bootstrap (it is much faster to approximate the score than to optimize the objective function). What is more, $\hat{\meat}_\bull(\btheta)$ is accurate for small bootstrap sample sizes (100 in our simulations). The procedure can be made even more efficient through parallelization.

\subsection{A two-stage semiparametric approach for continuous measurements}

If the sample size is large enough, a two-stage semiparametric method (SMP) may be used. In the first stage one estimates $F$ nonparametrically. The empirical distribution function $\hat{F}_n(y)=n^{-1}\sum_i 1\{Y_i\leq y\}$ is a natural choice for our estimator of $F$, but other sensible choices exist. For example, one might employ the Winsorized estimator
\begin{align*}
\tilde{F}_n(y)=\begin{cases}
\epsilon_n &\text{if }\,\hat{F}_n(y)<\epsilon_n\\
\hat{F}_n(y) &\text{if }\,\epsilon_n\leq\hat{F}_n(y)\leq 1-\epsilon_n\\
1-\epsilon_n &\text{if }\,\hat{F}_n(y)>1-\epsilon_n,
\end{cases}
\end{align*}
where $\epsilon_n$ is a truncation parameter \citep{klaassen1997efficient,liu2009nonparanormal}. A third possibility is a smoothed empirical distribution function
\[
\breve{F}_n(y)=\frac{1}{n}\sum_iK_n(y-Y_i),
\]
where $K_n$ is a kernel \citep{smoothedecdf}.

Armed with an estimate of $F$---$\hat{F}_n$, say---we compute $\hat{\bz}$, where $\hat{z}_i=\Phi^{-1}\{\hat{F}_n(y_i)\}$, and optimize
\begin{align*}
\ell_\textsc{ml}(\bomega\mid\hat{\bz})=-\frac{1}{2}\log\vert\momega\vert-\frac{1}{2}\hat{\bz}'\momega^{-1}\hat{\bz}
\end{align*}
to obtain $\hat{\bomega}$. This approach is advantageous when the marginal distribution is complicated, but has the drawback that uncertainty regarding the marginal distribution is not reflected in the (ML) estimate of $\hat{\bomega}$'s variance. This deficiency can be avoided by using a bootstrap sample $\{\hat{\bomega}^*_1,\dots,\hat{\bomega}^*_{n_b}\}$, the $j$th element of which can be generated by (1) simulating $\bU^*_j$ from the copula at $\bomega=\hat{\bomega}$; (2) computing a new response $\bY^*_j$ as $Y^*_{ji}=\hat{F}^{-1}_n(U^*_{ji})\;\;(i=1,\dots,n)$, where $\hat{F}^{-1}_n(p)$ is the empirical quantile function; and (3) applying the estimation procedure to $\bY^*_j$. We compute sample quantiles using the median-unbiased approach recommended by \citet{quantiles}. It is best to compute the bootstrap interval using the Gaussian method since that interval tends to have the desired coverage rate while the quantile method tends to produce an interval that is too narrow. This is because the upper-quantile estimator is inaccurate while the bootstrap estimator of $\var\hat{\bomega}$ is rather more accurate. To get adequate performance using the quantile method, a much larger sample size is required.

Although this approach may be necessary when the marginal distribution does not appear to take a familiar form, two-stage estimation does have a significant drawback, even for larger samples. If agreement is at least fair, dependence may be sufficient to pull the empirical marginal distribution away from the true marginal distribution. In such cases, simultaneous estimation of the marginal distribution and the copula should perform better. Development of such a method is beyond the scope of this article.

\section{Application to simulated data}
\label{simulation}

To investigate the performance of Sklar's $\omega$ relative to Krippendorff's $\alpha$, we applied both methods to simulated outcomes. We carried out a study for each level of measurement, for various sample sizes, and for a few values of $\omega$. The study plan is shown in Table~\ref{tab:simdes}, where $\mathcal{B}eta(\alpha, \beta)$ denotes a beta distribution, $\mathcal{L}(\mu,\sigma)$ denotes a Laplace distribution, $\phi(\mu,\sigma)$ denotes a Gaussian density function, $cat(\bp)$ denotes a categorical distribution, and $\mathcal{B}er(p)$ denotes a Bernoulli distribution.

\begin{table}[h]
   \centering
   \resizebox{\textwidth}{!}{
   \begin{tabular}{lccll}
   Margins & $\omega$ & \parbox[b]{1.75cm}{Sample\\Geometry\\\phantom{11.}$n_u,\,n_c$} & Method & Interval\\\hline
   $\mathcal{B}eta(1.5, 2)$\;\;\;\parbox[c][2.2em][b]{1cm}{\includegraphics[scale=.03]{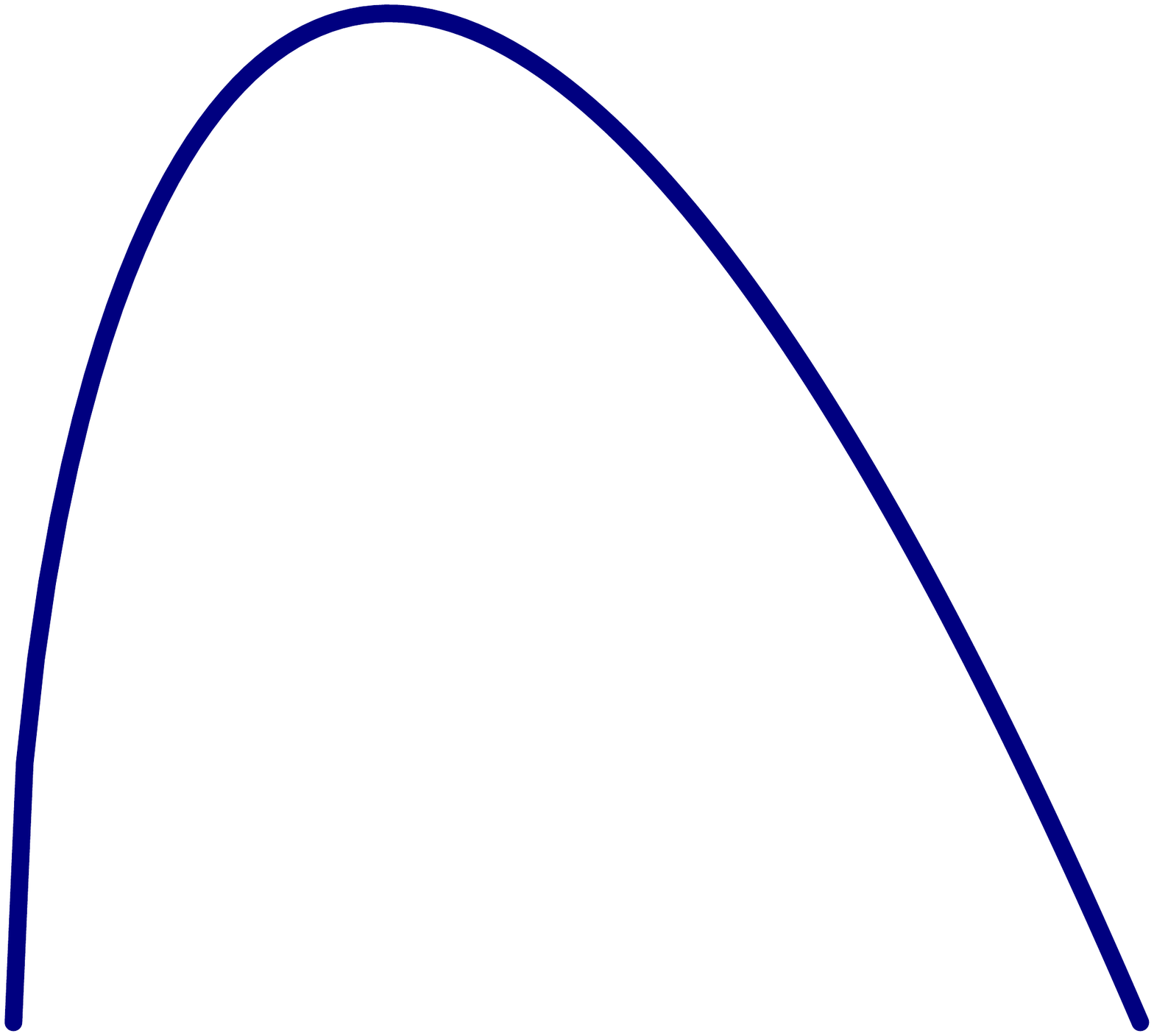}} & 0.70 & \phantom{1}30, \phantom{1}3 & ML &  ML\\
   $\mathcal{B}eta(13, 2)$\;\;\;\;\parbox[c][2.2em][b]{1cm}{\includegraphics[scale=.03]{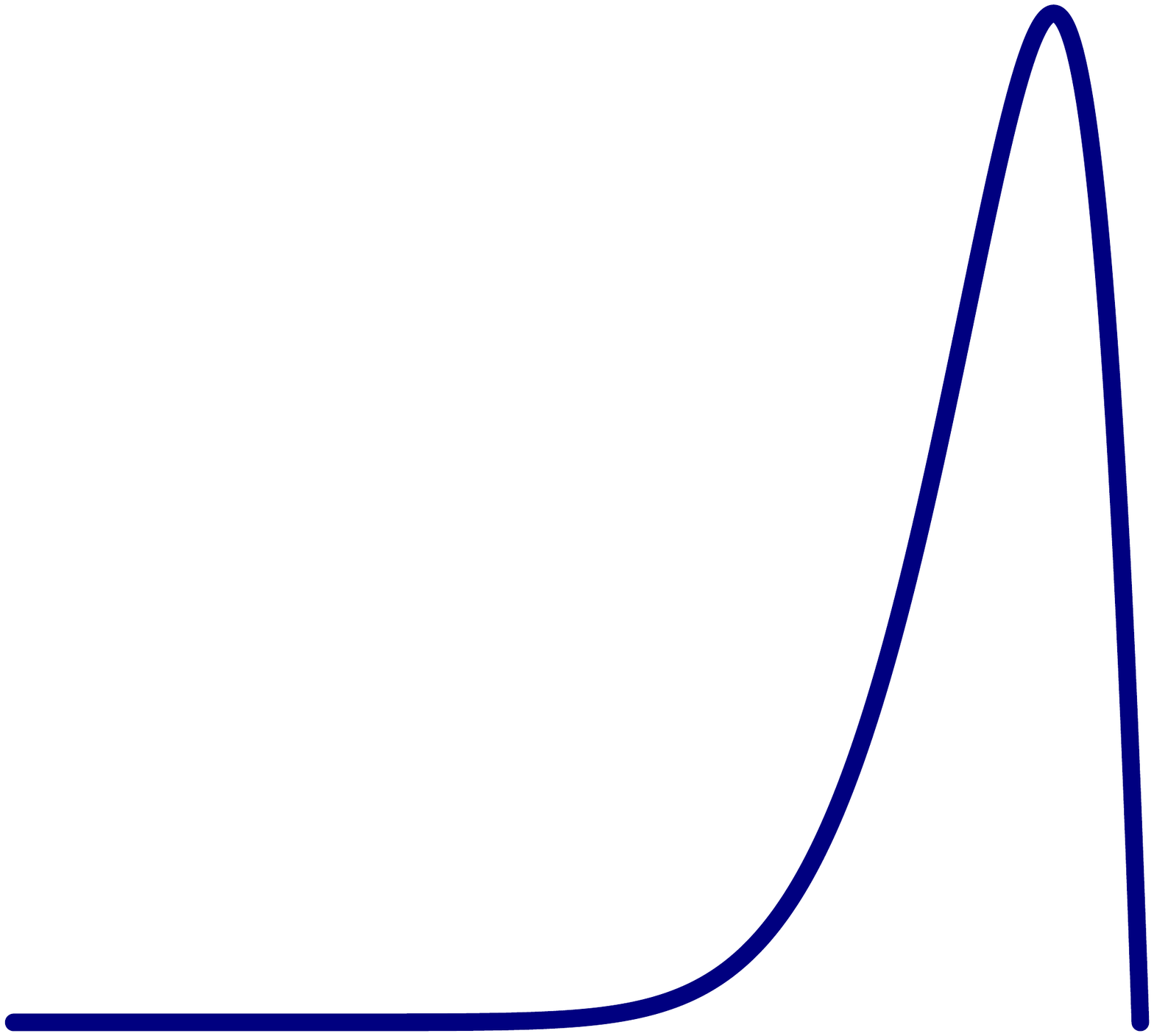}} & 0.95 & \phantom{1}10, \phantom{1}5 & ML & ML\\
   $\mathcal{L}(12, 4)$ & 0.65 & \phantom{1}40, \phantom{1}2 & ML & ML\\
   $0.3\,\phi(0, 1)+0.7\,\phi(3, 0.5)$\;\;\;\parbox[c][2.2em][b]{.6cm}{\includegraphics[scale=.03]{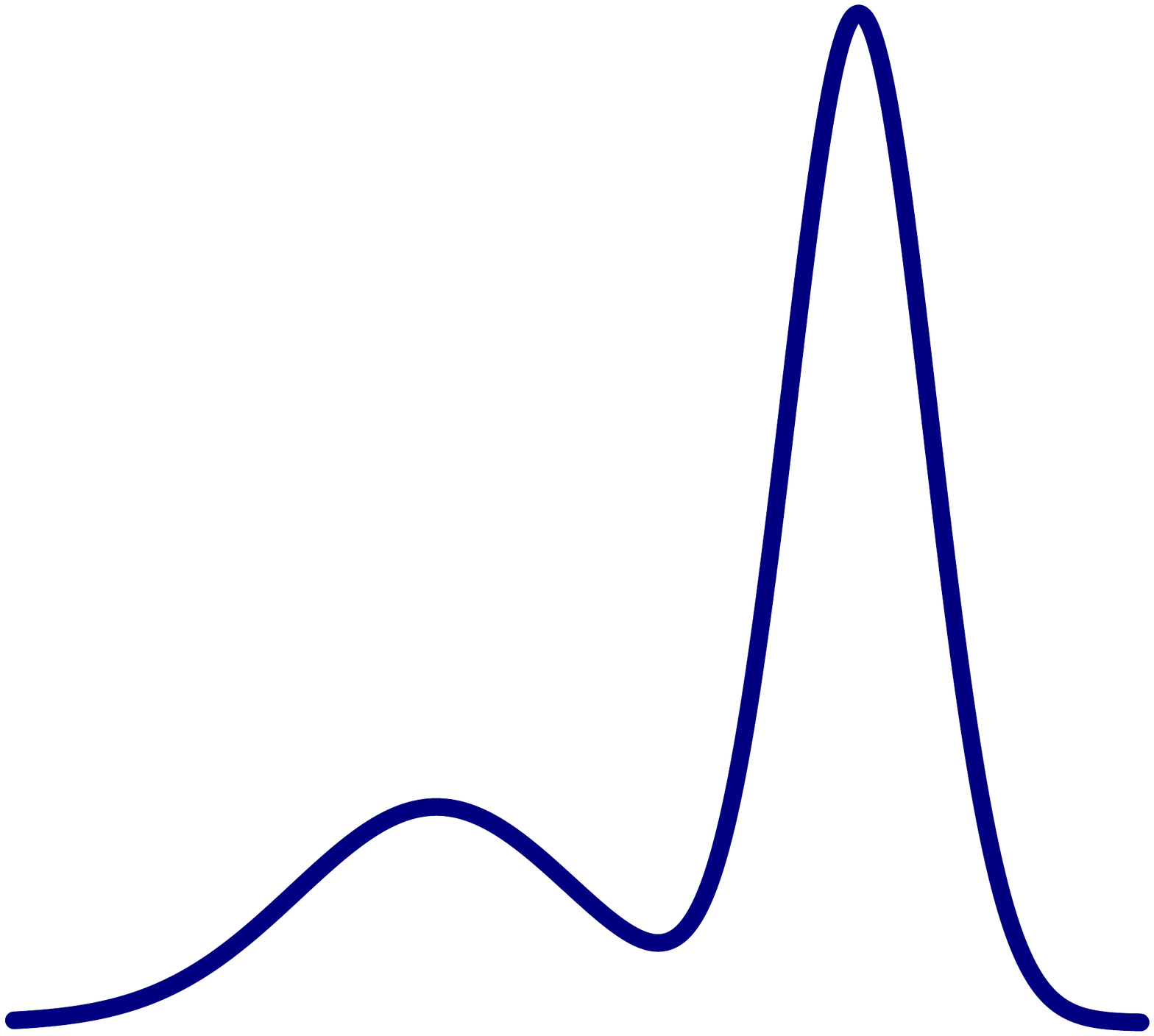}} & 0.80 & 100, \phantom{1}4 & SMP & Bootstrap\\
   $cat(0.1, 0.3, 0.2, 0.05, 0.35)$ & 0.90 & \phantom{1}20, 10 & DT & Sandwich\\
   $\mathcal{B}er(0.7)$ & 0.40 & 300, \phantom{1}6 & CML & Sandwich
    \end{tabular}}
    \caption{Our simulation scenarios. The images show the shapes of the beta and Gaussian-mixture densities.}
   \label{tab:simdes}
\end{table}

We applied Krippendorff's $\alpha$ and our procedure to each of 500--1,000 simulated datasets for each scenario. The results are shown in Table~\ref{tab:simres}. The coverage rates are for 95\% intervals. All of the intervals for $\alpha$ are bootstrap intervals.

For every simulation scenario, our estimator exhibited smaller bias, variance (excepting the final scenario), and mean squared error, and a significantly higher coverage rate. Our method proved especially advantageous for the first, fifth, and sixth scenarios. This is not surprising in light of the fact that Krippendorff's $\alpha$ implicitly assumes Gaussianity; the marginal distributions for the first, fifth, and sixth scenarios are far from Gaussian, and so Krippendorff's $\alpha$ performs relatively poorly for those scenarios. Krippendorff's $\alpha$ performed best for the second and third scenarios because the $\mathcal{B}eta(13, 2)$ and Laplace distributions do not depart too markedly from Gaussianity. Note that our estimator had a much larger variance than did $\hat{\alpha}$ for the final scenario. This is because we employed composite likelihood, which is a must for binary data since the DT estimator is badly biased for binary outcomes.

\begin{table}[h]
   \centering
   \resizebox{\textwidth}{!}{
   \begin{tabular}{lccrccc}
   Margins & $\omega$ & Median Est. & Bias & Variance & MSE & Coverage\\\hline
   \multirow{2}{*}{$\mathcal{B}eta(1.5, 2)$} & \multirow{2}{*}{0.70} & $\hat{\omega}=0.695$ & 2\% & 0.0065 & 0.0067 & 94\%\\
   & & $\hat{\alpha}=0.575$ & 19\% & 0.0073 & 0.0242 & 51\%\\\hline
   \multirow{2}{*}{$\mathcal{B}eta(13, 2)$} & \multirow{2}{*}{0.95} & $\hat{\omega}=0.942$ & 2\% & 0.0017 & 0.0021 & 95\%\\
   && $\hat{\alpha}=0.930$ & 3\% & 0.0021 & 0.0031 & 66\%\\\hline
   \multirow{2}{*}{$\mathcal{L}(12, 4)$} & \multirow{2}{*}{0.65} & $\hat{\omega}=0.651$ & 2\% & 0.0098 & 0.0099 & 93\%\\
   && $\hat{\alpha}=0.634$ & 4\% & 0.0111 & 0.0120 & 89\%\\\hline
   \multirow{2}{*}{$0.3\,\phi(0, 1)+0.7\,\phi(3, 0.5)$} & \multirow{2}{*}{0.80} & $\hat{\omega}=0.788$ & 2\% & 0.0008 & 0.0010 & 95\%\\
   && $\hat{\alpha}=0.756$ & 6\% & 0.0018 & 0.0040 & 73\%\\\hline
   \multirow{2}{*}{$cat(0.1, 0.3, 0.2, 0.05, 0.35)$} & \multirow{2}{*}{0.90} & $\hat{\omega}=0.900$ & $<$ 1\% & 0.0010 & 0.0010 & 98\%\\
   && $\hat{\alpha}=0.504$ & 44\% & 0.0052 & 0.1626 & \phantom{8}0\%\\\hline
   \multirow{2}{*}{$\mathcal{B}er(0.7)$} & \multirow{2}{*}{0.40} & $\hat{\omega}=0.397$ & 6\% & 0.0173 & 0.0180 & 93\%\\
   && $\hat{\alpha}=0.244$ & 38\% & 0.0007 & 0.0240 & \phantom{8}0\%
    \end{tabular}}
    \caption{Results from our simulation study.}
   \label{tab:simres}
\end{table}

\section{R package \texttt{sklarsomega}}
\label{package}

Here we briefly introduce our R package, \texttt{sklarsomega}, by way of a usage example. The package is available for download from the Comprehensive R Archive Network.

The following example applies Sklar's $\omega$ to the nominal data from the first case study of Section~\ref{examples}. We provide the value \texttt{"asymptotic"} for argument \texttt{confint}. This results in sandwich estimation for the DT approach (see Section~\ref{sandwich}). We estimate $\meat$ using a parallel bootstrap, where $n_b=$ 1,000 and six CPU cores are employed (control parameter \texttt{type} takes the default value of \texttt{"SOCK"}). Since argument \texttt{verbose} was set to \texttt{TRUE}, a progress bar appears \citep{pbapply}. We see that $\hat{\meat}$ was computed in one minute and 49 seconds.

\begin{scriptsize}
\begin{verbatim}
> data = matrix(c(1,2,3,3,2,1,4,1,2,NA,NA,NA,
+                 1,2,3,3,2,2,4,1,2,5,NA,3,
+                 NA,3,3,3,2,3,4,2,2,5,1,NA,
+                 1,2,3,3,2,4,4,1,2,5,1,NA), 12, 4)
> colnames(data) = c("c.1.1", "c.2.1", "c.3.1", "c.4.1")
> fit = sklars.omega(data, level = "nominal", confint = "asymptotic", verbose = TRUE,
+                    control = list(bootit = 1000, parallel = TRUE, nodes = 6))

Control parameter 'type' must be "SOCK", "PVM", "MPI", or "NWS". Setting it to "SOCK".

   |++++++++++++++++++++++++++++++++++++++++++++++++++| 100% elapsed = 01m 49s
> summary(fit)

Call:

sklars.omega(data = data, level = "nominal", confint = "asymptotic", 
    verbose = TRUE, control = list(bootit = 1000, parallel = TRUE, 
        nodes = 6))

Convergence:

Optimization converged at -40.42 after 31 iterations.

Control parameters:
                    
bootit   1000       
parallel TRUE       
nodes    6          
dist     categorical
type     SOCK       
                    
Coefficients:

      Estimate    Lower  Upper
inter  0.89420  0.76570 1.0230
p1     0.25170  0.01407 0.4893
p2     0.24070  0.01842 0.4631
p3     0.22740  0.04639 0.4084
p4     0.18880 -0.06007 0.4377
p5     0.09136 -0.15580 0.3385
\end{verbatim}
\end{scriptsize}

Next we compute DFBETAs \citep{Young2017Handbook-of-Reg} for units 6 and 11, and for coders 2 and 3. We see that omitting unit 6 results in a much larger value for $\hat{\omega}$, whereas unit 11 is not influential. Likewise, coder 2 is influential while coder 3 is not.

\begin{scriptsize}
\begin{verbatim}
> (inf = influence(fit, units = c(6, 11), coders = c(2, 3)))
$dfbeta.units
         inter         p1           p2          p3          p4           p5
6  -0.07914843 0.03438538  0.052599491 -0.05540904 -0.05820757  0.026631732
11  0.01096758 0.04546670 -0.007630807 -0.01626192 -0.01514173 -0.006432246

$dfbeta.coders
          inter           p1           p2          p3         p4          p5
2  0.0579843781 -0.002743713  0.002974195 -0.02730064 0.01105672  0.01601343
3 -0.0008664934 -0.006572821 -0.048168128  0.05659853 0.02149364 -0.02335122

> fit$coef - t(inf$dfbeta.units)
               6         11
inter 0.97335265 0.88323664
p1    0.21731494 0.20623362
p2    0.18814896 0.24837926
p3    0.28280614 0.24365903
p4    0.24700331 0.20393747
p5    0.06472664 0.09779062
\end{verbatim}
\end{scriptsize}

Much additional functionality is supported by \texttt{sklarsomega}, e.g., plotting, simulation. And we note that computational efficiency is supported by our use of sparse-matrix routines \citep{Furrer:Sain:2010:JSSOBK:v36i10} and a clever bit of Fortran code \citep{genz1992numerical}. Future versions of the package will employ C++ \citep{Eddelbuettel:Francois:2011:JSSOBK:v40i08}.

\section{Conclusion}
\label{conclusion}

Sklar's $\omega$ offers a flexible, principled, complete framework for doing statistical inference regarding agreement. In this article we developed various frequentist approaches for Sklar's $\omega$, namely, maximum likelihood, distributional-transform approximation, composite marginal likelihood, and a two-stage semiparametric method. This was necessary because a single, unified approach does not exist for the form of Sklar's $\omega$ presented in Section~\ref{method}, wherein the copula is applied directly to the outcomes. An appealing alternative would be a hierarchical formulation of Sklar's $\omega$ such that the copula is applied through the responses' mean structure. This would permit, for example, a well-motivated Bayesian scheme and/or expectation-maximization algorithm to be devised.

Another potentially appealing extension/refinement would focus on composite likelihood inference for the current version of the model. Perhaps one should use a different composite likelihood, for example, and/or employ well-chosen weights \citep{xu2016note}.

\bibliography{refs}
\bibliographystyle{apa}

\end{document}